\documentclass[aip,jcp,reprint, amsmath,amssymb]{revtex4-1}
\usepackage[usenames]{color}
\usepackage{graphicx}
\usepackage{dcolumn}
\usepackage{bm}
\usepackage{ulem}
\usepackage{amsmath}
\usepackage{amssymb}

\begin{document}

\title{Rotational Tunneling States and the Non-Debye Specific Heat of Dipolar Glasses}

\author{P.S. Goyal}
\affiliation{Pillai's Institute for Information Technology, Engineering, Media Studies \& Research, New Panvel, Mumbai -410206, India}
\author{P.D. Babu}
\affiliation{UGC-DAE Consortium for Scientific Research, Mumbai Centre, R5 Shed, BARC, Mumbai-400085, India}
\author{F. Juranyi}
\affiliation{Laboratry for Neutron Scattering, Paul Scherrer Institut, 5232 Villigen, Switzerland}

\begin{abstract}

Specific heat of dipolar glasses does not obey Debye law. It is of interest to know if the non-Debye specific heat can be accounted for in terms of Schottky-type specific heat arising from rotational tunneling states of the dipoles. This paper deals with rotational tunneling spectra of NH$_{4}^{+}$ ions and the non-Debye specific heat of mixed salts (e.g. (NH$_{4})_{x}$Rb$_{1-x}$Br) of ammonium and alkali halides which are known to exhibit dipolar glass phase. We have measured specific heat of above mixed salts at low temperatures (1.5 K $< T <$ 15 K). It is seen that while the specific heat of pure salts obeys Debye law, the specific heat of mixed salts does not obey Debye law. We have studied the effect of the NH$_{4}^{+}$ ion concentration, first neighbor environment of NH$_{4}^{+}$ ion and the lattice strain field on the non-Debye specific heat by carrying out measurements on suitably chosen mixed salts. Independent of above, we have measured the rotational tunneling spectra, $f(\omega $), of the NH$_{4}$ ions in above salts using technique of neutron incoherent inelastic scattering. The above studies show that both the non-Debye specific heat and the tunneling spectra of the NH$_{4}^{+}$ ions depend on the NH$_{4}^{+}$ ion concentration, first neighbor environment of NH$_{4}^{+}$ ions and the lattice strain field. We have further shown that the temperature dependence of the measured specific heat can be explained for all the samples in terms of a model that takes account of contributions to the specific heat from the Debye phonons and the rotational tunneling states of the NH$_{4}^{+}$ ions. To the best of our knowledge, this is a first study where it is shown that measured specific heat of (NH$_{4})_{x}$Rb$_{1-x}$Br can be quantitatively explained in terms of an experimentally measured rotational tunneling spectra $f(\omega $) of the NH$_{4}^{+}$ ions.

\end{abstract}



\maketitle

\section{INTRODUCTION}

Dipolar glasses are crystalline materials in which electric dipoles are frozen in random directions.\cite{r1,r2,r3,r4,r5,r6,r7} Specific heat of such dipolar glasses does not obey Debye law ($C \sim T^{3}$ at low temperatures). It is known for quite some time that non-Debye specific heat of dipolar glasses is connected with the Schottky specific heat that arises from rotational tunneling states of the dipoles.\cite{r8,r9} However, so far it has not been possible to quantitatively explain the measured specific heat in terms of tunneling spectra $f$($\omega $) which is independently measured using a spectroscopic technique. This paper deals with measurement of rotational tunneling spectra $f$($\omega $) of NH$_{4}^{+}$ ions in mixed salts (NH$_{4}$)$_{x}$Rb$_{1-x}$Br of ammonium and rubidium bromides and correlating it with the specific heat of above salts at low temperatures (T $<$ 15 K). The rotational tunneling spectrum $f$($\omega $) of NH$_{4}^{+}$ ions has been measured using neutron incoherent inelastic scattering.

Dipolar glasses consist of materials, some of whose lattice sites are occupied by molecules or molecular groups having finite electric dipole moment. These dipoles have orientation degrees of freedom and below some freezing temperature T$_{f}$, they freeze in random directions. That is, unlike ``window'' glasses where the frozen-in disorder is translational, the frozen-in disorder is rotational in dipolar glasses.\cite{r1} K(CN)$_{x}$Br$_{1-x}$ is an example of a dipolar glass where CN$^{-}$ ions have electric dipole moments and they occupy some of the Br sites of KBr lattice.\cite{r1,r2} (NH$_{4}$)$_{x}$K$_{1-x}$I is an other example of a dipolar glass where NH$_{4}^{+}$ ions have electric dipole moments and they occupy some of the K$^{+}$ sites of KI lattice.\cite{r5,r6}  It may be mentioned that in general, NH$_{4}^{+}$ ion does not possess electric dipole moment as it is tetrahedron in shape (spherical symmetric). However, NH$_{4}^{+}$ ion gets distorted and develops electric dipole moment in mixed salts (NH$_{4}$)$_{x}$A$_{1-x}$Y (A= Rb, K; Y = Br, I)  of ammonium and alkali halides.

The motion of a dipole or a linear molecule such as CN$^{-}$ in a dipolar glass depends on the orientational potential V($\theta $) experienced by the dipole.\cite{r10,r11} Usually the dipole undergoes harmonic oscillatory motion about its mean orientation. However if the hindering potential is weak, the dipoles undergo reorientational motion (i.e. large angular jump) over and above the oscillatory motion. The reorientational motion, which involves reorientation of the dipole from one orientation to another, is a thermally activated process. The above stochastic reorientation process exhibits as a quantum mechanical rotational tunneling process at low temperatures.\cite{r10,r11}  The rotational tunneling gives rise to the splitting of the rotational ground state of the dipole and the magnitude of splitting energy (referred to as rotational tunneling energy) depends on the strength of the hindering potential V($\theta $). This splitting of ground state has been seen for K(CN)$_{x}$Br$_{1-x}$ , where CN$^{-}$ ions reorient by 180$^\circ$ via quantum mechanical tunneling at low temperatures.\cite{r8,r9} Neutron incoherent inelastic scattering is an excellent technique for measuring rotational tunneling states in solids. The subject of tunneling states and their study using neutrons has been discussed at great length in the book by Press\cite{r10} and in the review article by Prager and Heidemann.\cite{r11}

The mixed salts (NH$_{4}$)$_{x}$A$_{1-x}$Y (A= Rb, K; Y = Br, I) of ammonium and alkali halides have NaCl structure down to low temperatures.\cite{r12,r13,r14,r15}  The NH$_{4}^{+}$ ion in above salts resides in an octahedral environment of first-neighbor halide ions and the rotational hindering potential V($\theta $) experienced by the NH$_{4}^{+}$ ions is quite weak. There are several publications dealing with the tunneling states of NH$_{4}^{+}$ ions\cite{r6,r7,r15,r16,r17,r18,r19,r20,r21} of mixed salts of ammonium and alkali halides, including a review paper by David Smith.\cite{r22}  The energies associated with rotational tunneling spectra of NH$_{4}^{+}$ ions are typically in range of 0.0 to 0.7 meV depending on the composition of the mixed salt.\cite{r11,r22} There are several publications dealing with the specific heat of mixed salts of ammonium and alkali halides also.\cite{r16,r17,r18,r23} It is seen that the specific heat of above salts does not obey Debye law and the details of non-Debye specific heat depend on the composition of the mixed salt. There are, however, no systematic studies which bring out the role of first neighbor halide ion Y, the NH$_{4}^{+}$ ion concentration $x$ and the lattice strain (arising from the mismatch of sizes of cations) on the tunneling spectra and the non-Debye specific heat of above salts. More importantly, it is not clear if the non-Debye specific heat of above salts can be quantitatively accounted for in terms of the rotational tunneling spectra of the NH$_{4}^{+}$ ions. Among the well known dipolar glasses, the mixed salts of ammonium and alkali halides provide an ideal system for such a study as the rotational tunneling spectra of NH$_{4}^{+}$ ion can be easily measured using neutrons. This is connected with the fact that hydrogen has high cross-section for neutron incoherent scattering.\cite{r10,r24}  It may be mentioned that Loidl \textit{et.al.}\cite{r9} had measured rotational tunneling spectra of CN$^{-}$ ions in K(CN)$_{x}$Br$_{1-x}$ using neutrons but the quality of data was not good as neutron incoherent scattering from non-hydrogenous CN$^{-}$ ions is quite weak.

In view of above, we have carried out a fresh set of incoherent neutron inelastic scattering (INIS) and specific heat measurements on the mixed salts of ammonium and alkali halides. The role of NH$_{4}^{+}$ concentration on the tunneling spectra of NH$_{4}^{+}$ ions has been examined by measuring the rotational tunneling spectra of the NH$_{4}^{+}$ ions in (NH$_{4})_{x}$Rb$_{1-x}$Br for several values of $x$. Similarly, we have carried out the comparative studies on (NH$_{4})_{0.02}$K$_{0.98}$I and (NH$_{4})_{0.02}$K$_{0.98}$Br for getting an insight on the role of first neighbor halide ion on the tunneling spectra of the NH$_{4}^{+}$ ions. Further, the effect of strain field on tunneling spectra of NH$_{4}$ ions has been examined by carrying out neutron measurements on ternary salts (NH$_{4})_{0.02}(Rb_{y}$K$_{1-y}$)$_{0.98}$Br. The above ternary salts provide a system where one can change the lattice strain without altering the NH$_{4}^{+}$ concentration; this is achieved by changing the relative concentration y of K$^{+}$ and Rb$^{+}$ ions, whose sizes are different. In addition to above, we have measured specific heat for (NH$_{4}$)$_{x}$Rb$_{1-x}$Br and (NH$_{4})_{0.02}(Rb_{y}$K$_{1-y})_{0.98}$Br series for temperatures down to 1.5 K. We have calculated the specific heat of above salts using the measured tunneling spectra $f$($\omega $) of the NH$_{4}^{+}$ ions and compared the calculated specific heat with the measured one. Both neutron and specific heat measurements were made on the same sample so that there are no errors arising because of uncertainty in NH$_{4}^{+}$ ions concentration etc. It is seen that temperature dependence of measured specific heat can be explained for all the samples in terms of a model that takes account of contributions to the specific heat from Debye phonons and the rotational tunneling states of the NH$_{4}^{+ }$ions.

The plan of the paper is as follows. The next section gives experimental details. Results of incoherent neutron inelastic scattering measurements are given in Section 3 and those of specific heat measurements are given in Section 4. Section 5 deals with the calculation of specific heat based on neutron data, and a comparison between the calculated and the measured specific heats. In the end, Section 6 gives a summary.

\section{EXPERIMENTAL DETAILS}

\subsection{Sample Preparation and Characterization}

The mixed salts (NH$_{4}$)$_{x}$Rb$_{1-x}$Br and (NH$_{4}$)$_{x}$K$_{1-x}$Br for $x$ = 0.02, 0.05 and 0.10 (now onwards referred to as NRB02, NRB05, NRB10 and NKB02, NKB05, NKB10), (NH$_{4}$)$_{0.02}$K$_{0.98}$I and (NH$_{4})_{0.02}($Rb$_{0.5}$K$_{0.5})_{0.98}$Br (referred to as NKI02 and NKRB02) were prepared by re-crystallization from an aqueous solution containing appropriate amounts of pure salts.\cite{r12,r14} The starting chemicals NH$_{4}$Br, NH$_{4}$I, RbBr, RbI, KBr and KI with 99.99\% purity were procured from Aldrich. The re-crystallization of the mixed salts were carried out at a fixed temperature of 50$^\circ$ C. The powder samples of the mixed salts were characterized using X-ray and neutron diffraction. X-ray studies confirmed that samples have crystallized in NaCl structure and are of single phase nature. Lattice constant of mixed salts (NH$_{4})_{x}$A$_{1-x}$Y were found to increase with $x$ as expected from Vegard law.\cite{r25} A typical neutron diffraction peak of (200) for (NH$_{4}$)$_{x}$Rb$_{1-x}$Br corresponding to $x$ = 0.02 and $x$ = 0.10 is shown in Fig.1. It is seen that lattice constant of (NH$_{4})_{x}$Rb$_{1-x}$Br increases by about 0.2 \% when $x$ is increased from 0.02 to 0.10. The low temperature neutron diffraction experiments, including the present ones, showed that above mixed salts retain NaCl structure down to the lowest temperature.\cite{r15,r19}

\begin{figure}
  \centering
   \includegraphics[width=3.5in,trim=0.09in 0.30in 0.09in 0.13in]{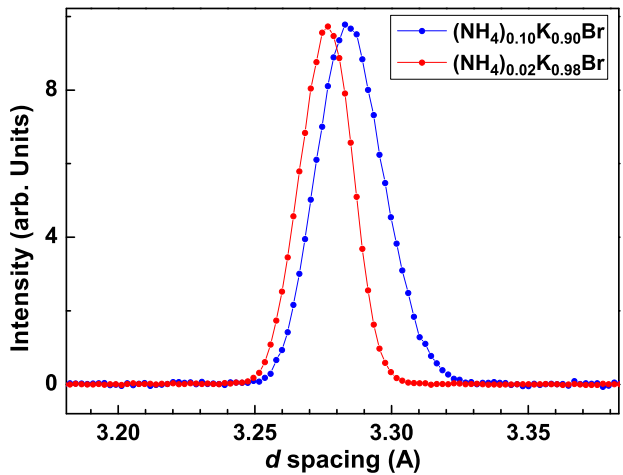}\\
  \caption{Diffraction peak (200)for mixed salt (NH$_{4})_{x}$K$_{1-x}$Br for x = 0.02 and 0.10 as recorded at 1.8 K using diffraction detectors of MARS. The lattice spacing increases by about 0.2\% when NH$_{4}^{+}$ concentration is increased from 0.02 to 0.10.}\label{fig1}
\end{figure}

\subsection{Incoherent Neutron Inelastic Scattering}

We have carried out incoherent neutron inelastic scattering (INIS) experiments on above samples using MARS spectrometer at spallation neutron facility SINQ, Paul Scherrer Institut, Switzerland. The above spectrometer is an inverted geometry spectrometer where incident energy E$_{i}$ is analyzed using time-of-flight technique and the final energy is kept fixed at E$_{f}$ = 1.87 meV using mica analyzers in the back scattering position.\cite{r26} The spectrometer provides an energy resolution of $\sim$ 13 $\mu$eV at zero energy transfer ($\hbar \omega $ = 0). The measurements were made both, on mixed salts (NRB02, NRB05, NRB10; NKB02, NKI02, and NKRB02) and dummy samples (RbBr, RbI, KBr and KI) for an energy transfer $\hbar\omega$ (= E$_{i}$ - E$_{f}$) range of -0.2 meV to +0.70 meV.\cite{r27} The sample temperature was kept at 1.8 K in all the measurements. The sample holder was a cylinder made of Aluminium and having inner diameter of 10 mm. The data were compared with a standard Vanadium sample, and it was seen that scattering was less than 10\% for all the samples.

It may be mentioned that the time of flight experiment of the type described above measures scattering law S$_{inc}(\Phi, \omega$) for the specimen.\cite{r28} We have measured S$_{inc}(\Phi, \omega )$ for our samples corresponding to five scattering angles ($\Phi = 30^\circ, 60^\circ, 90^\circ, 120^\circ$ and $150^\circ$) using MARS. The structural information about the samples was also obtained simultaneously using time of flight diffraction detectors of MARS.

\subsection{Specific Heat Measurements}

We have measured specific heat of above salts by the thermal relaxation method using the heat capacity option at PPMS (Quantum Design, USA). The sample used was about 10-20 mg and it was fixed on sample holder using N-Apiezon grease. The sample was first cooled to the lowest temperature of measurement and the specific heat data were recorded during warming cycle. The heat capacity of the sample is obtained by subtracting the previously measured addenda (empty sample holder + N grease). The specific heat measurements have been made on mixed salts NRB02, NRB05, NRB10, NKB02 and NKRB02 and the corresponding pure salts (NH$_{4}$Br, KBr, RbBr and Rb$_{0.5}$K$_{0.5}$Br) for sample temperatures between 1.5K and 15K.\cite{r29}

\section{RESULTS AND DISCUSSIONS: NEUTRON EXPERIMENTS}

\subsection {Effect of Change in First Neighbor Halide Ion on the Tunneling Spectra}

We have carried out the comparative neutron studies on (NH$_{4})_{0.02}$K$_{0.98}$I and (NH$_{4})_{0.02}$K$_{0.98}$Br for examining the effect of change in first neighbor halide ion on the tunneling spectra of the NH$_{4}^{+}$ ions. The measured S$_{inc}(\Phi, \omega $) for the two salts are shown in Fig.2. The data have been corrected for contributions from the sample holder and for the background neutrons, which are scattered from the sample directly into the detector. The two distributions have been normalized for a fixed elastic intensity and thus they correspond to a fixed number of NH$_{4}^{+}$ ions. It may be mentioned that the distributions shown in Fig.2 correspond to a mean scattering angle of $\Phi = 90^\circ $ as the data from all the detectors have been added. It is seen that there is a strong peak at about 0.606 meV for NKI02 and the corresponding peak for NKB02 is seen at $\sim$ 0.525 meV. The width of above peak is 27 $\mu$eV for NKI02 and 56 $\mu $eV for NKB02. In addition to this, there is a broad distribution around $h\omega = 0$ for NKB02 which is not seen for NKI02.  The width of this broad component under elastic peak is ($\sim 180 \mu$eV) is much more than the instrumental resolution (13 $\mu$eV).

\begin{figure}
  \centering
   \includegraphics[width=3.4in,trim=0.09in 0.30in 0.09in 0.13in]{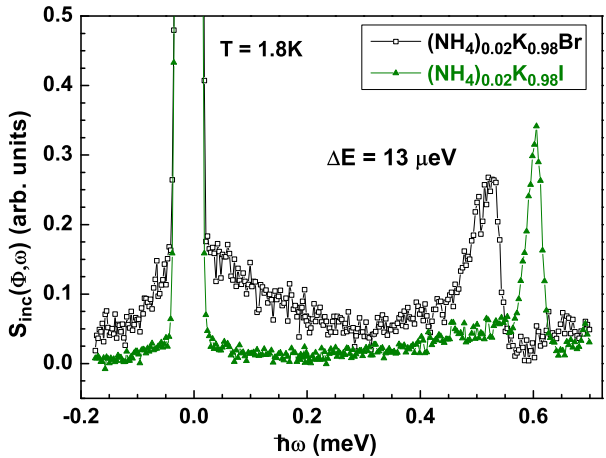}\\
  \caption{The variation of measured S$_{inc}(\Phi,\omega $) with $\omega $ for NKI02 and NKB02 at 1.8K.  It is seen that tunneling spectrum of NH$_{4}^{+}$ ion depends on the first neighbor halide ion.}\label{fig2}
\end{figure}

In the following, we discuss the origin of sharp peaks and postpone the discussion on the broad feature to the next section (III B). Neutron incoherent scattering from above salts is mainly from protons of NH$_{4}^{+}$ ions and thus the peaks in the measured S$_{inc}(\Phi,\omega $) arise from the various vibration modes (translational, rotational, internal vibrations or tunneling) of NH$_{4}^{+}$ ions.\cite{r10,r24} However, Neutron\cite{r30,r31,r32} IR\cite{r13} and Raman scattering\cite{r33} studies for NaCl phase of ammonium halides show that energies associated with the vibration modes (translational optical, rotational or internal vibrations) of NH$_{4}^{+}$ ions are quite large ($>$ 5 meV). Moreover, David Smith\cite{r22} has reviewed the experimental results on the rotational motion of NH$_{4}^{+}$ ions in a number of ammonium compounds and he has shown that the energy associated with rotational tunneling modes of NH$_{4}^{+}$ ions is $\sim $ 0.5 meV, and the tunneling region is well separated from the energy region of rotational modes ( $\sim$ 20 meV). In view of above, we believe that vibration modes do not contribute in energy region of -0.2 to +0.70 meV and the sharp peaks seen in Fig.2 are connected with the rotational tunneling states of NH$_{4}^{+}$ ions. That is, peak at 0.606 meV for NKI02 and that at 0.525 meV for NKB02 in measured S$_{inc}(\Phi,\omega $) arise because of the transitions between the rotational tunneling states of NH$_{4}$ ions.\cite{r10,r22} The value of 0.606 meV for tunneling energy of NH$_{4}^{+}$ ions in NKI02 is in excellent agreement with the earlier studies.\cite{r6, r15,r17,r19,r22}  Similarly, the present tunneling energy value of 0.525 meV for NKB02 is also reasonable as earlier studies showed that tunneling energy is 0.530 meV for x = 0.005\cite{r17} and 0.516 meV for x = 0.032.\cite{r19} In view of the differences in the peak positions, the peak widths, and the existence or non-existence of broad distribution underneath the elastic peak, we conclude that the rotational tunneling spectra of NH$_{4}^{+}$ ions in NKB02 are quite different from those in NKI02. It is clear from above studies that a change in first neighbor halide ion modifies the hindering potential V($\theta $) and that in turn changes the tunneling spectrum of the NH$_{4}^{+}$ ions in the mixed salts of ammonium alkali halides. In principle, the tunneling data can be used to get details of the hindering potential V($\theta $) experienced by NH$_{4}^{+}$ ion.\cite{r22,r34,r35} We are, however, largely concerned with establishing a correlation between the tunneling spectra and non-Debye specific heat and will thus not get into the calculation of potentials etc.
\subsection{Effect of NH$_{4}^{+}$ Concentration on Tunneling Spectra}

The effect of NH$_{4}^{+}$ ion concentration on the measured S$_{inc}$($\Phi $, $\omega $) for (NH$_{4}$)$_{x}$Rb$_{1-x}$Br are shown in Fig.3. The three distributions corresponding to NRB02 (x = 0.02), NRB05 (x = 0.05) and NRB10 (x = 0.10) have been normalized for a fixed elastic intensity and they correspond to a fixed number of NH$_{4}^{+}$ ions. It is seen that S$_{inc}$($\Phi $, $\omega $) has a strong peak at about 0.46 meV and a small peak at about 0.34 meV for all the three samples. The above peaks correspond to the tunneling energies of NH$_{4}^{+}$ ions in these salts. The present data are in agreement with earlier studies of Mukhopadhyay \textit{et al.}\cite{r7} who found that scattered neutron spectrum from (NH$_{4}$)$_{x}$Rb$_{1-x}$Br for x = 0.03 has peaks at about 0.47 meV and 0.35 meV.

Figure 3 shows that, in addition to the well defined peaks, the scattered neutron spectrum has a broad continuous distribution centered around $\hbar \omega $ = 0. As compared to instrument resolution of $\sim$ 13 ľeV, the width of the broad distribution for the three samples are 60 $\mu $eV, 90 $\mu $eV and 130 $\mu $eV for $x$ = 0.02, 0.05 and 0.10, respectively. Even though the positions of tunneling peaks (0.46 meV and 0.34 meV), are independent of $x$, the intensity of the broad distribution increases with increase in NH$_{4}^{+}$ ion concentration. It may be mentioned that usually broad distribution around $\hbar \omega  = 0$ arises because of thermally activated stochastic dynamical processes (such as random molecular reorientations) and is referred to as quasi-elastic scattering in the literature.\cite{r10} However, the broad distribution seen in Fig.3 can not be connected with the stochastic processes as the sample temperature is very low ($\sim$ 1.8K) and the random dynamical processes are expected to be very slow at these temperatures. It is thus believed that broad distribution in S$_{inc}$($\Phi $, $\omega $)  is connected with a distribution of tunneling energies. We believe that the broad distribution seen for NKB02 (Fig.2) is also connected with the existence of a distribution of low energy tunneling states in this salt. Colmenero \textit{et al.}\cite{r36} had also observed that a distribution in tunneling energies gives rise to a broad intensity distribution underneath the elastic peak in neutron experiments.

\begin{figure}
  \centering
   \includegraphics[width=3.4in,trim=0.09in 0.30in 0.09in 0.13in]{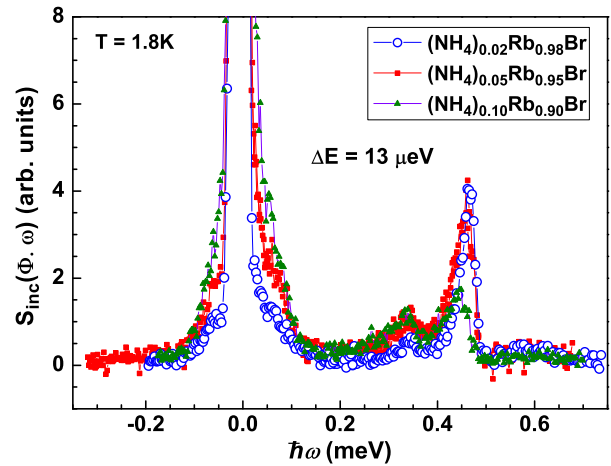}\\
  \caption{Effect of NH$_{4}^{+}$concentration on the measured inelastic neutron spectra for (NH$_{4})_{x}$Rb$_{1-x}$Br at 1.8 K.  The intensity of broad distribution around $h\omega = 0$ increases with increase in NH$_{4}^{+}$ concentration.}\label{fig3}
\end{figure}

The existence of more than one tunneling peak and the distribution of tunneling energies near $\hbar \omega $ = 0 in Fig.3 suggests that all NH$_{4}^{+}$ ions in (NH$_{4})_{x}$Rb$_{1-x}$Br do not have same environment.\cite{r6,r21} In general, NH$_{4}^{+}$ ion in (NH$_{4})_{x}$Rb$_{1-x}$Br has halide as the first neighbor and cation (Rb or NH$_{4}^{+}$ ion) as second neighbor. Ideally, the number ($n$) of 2nd neighbor NH$_{4}^{+}$ ion should be same for all the  NH$_{4}^{+}$ ions. In case of non-uniform distribution of NH$_{4}^{+}$ ions, the number of second neighbor NH$_{4}^{+}$ ions seen by an ammonium ion could be different (say $n$ = 0, 1, 2, ... etc.) at different sites. This results in a distribution of potentials and a distribution of tunneling energies. It seems that the main peak at 0.46 meV corresponds to an NH$_{4}^{+}$ ion for which all twelve cation sites are occupied by Rb ions and there is no ammonium ion ($n$ = 0) at those sites. Similarly, the peak at 0.34 meV could corresponds to an NH$_{4}^{+}$ ion for which 11 cation sites are occupied by Rb ions and one site is occupied ammonium ion ($n$ = 1). The lower tunneling energy for $n$ = 1 as compared to that for $n$ = 0 is connected with the fact that NH$_{4}^{+}$ - NH$_{4}^{+}$ interaction leads to a stronger potential V($\theta $) for $n$ =1. An increase in NH$_{4}^{+}$ concentration $x$ results in an increase in the number $n$ of neighboring ammonium ions, which in turn increases the strength of the hindering potential. The increase in intensity of broad distribution of tunneling energies near $\hbar \omega $ = 0 at the cost of intensity of isolated tunneling peak, is a reflection of the fact that strength of potential experienced by  NH$_{4}^{+}$ ion increases with increase in  NH$_{4}^{+}$ ion concentration. It may be noted that increase in strength of the hindering potential is partly because of increase in  NH$_{4}^{+}$ -  NH$_{4}^{+}$ interaction and partly because of increase in the strain field (arising because of mismatch of ionic sizes) in the lattice. The strain field, in fact, mimics the effect of NH$_{4}^{+}$ - NH$_{4}^{+}$ interaction in increasing the strength of hindering potential (See next section)

\subsection{Effect of Lattice Strain on Tunneling Spectra}

The sizes of NH$_{4}^{+}$ ion (1.48{\AA}), Rb$^{+}$ (1.47{\AA}) ion and K$^{+}$ (1.33{\AA}) ion are different and as a result of that, the lattice of mixed salt is strained. The actual value of strain depends on the composition of the mixed salt. For example, it is expected that the lattice strain in (NH$_{4}$)$_{x}$K$_{1-x}$Br would be more than that in (NH$_{4}$)$_{x}$Rb$_{1-x}$Br. A change in lattice strain changes the hindering potential V ($\theta $) and that in turn modifies the tunneling spectrum of NH$_{4}^{+}$ ion. We believe that differences, though small, in tunneling spectra of NH$_{4}^{+}$ ions in (NH$_{4}$)$_{0.02}$K$_{0.98}$Br (Fig.2) and (NH$_{4})_{0.02}$Rb$_{0.98}$Br (Fig.3) are connected with the differences in their lattice strains. This section deals with a study of the effect of the lattice strain on the tunneling spectrum of NH$_{4}^{+}$ ions in mixed salts of ammonium and alkali halides.

The strain in (NH$_{4}$)$_{x}$Rb$_{1-x}$Br lattice increases with increase in $x$ because of differences in sizes of NH$_{4}^{+}$ and Rb$^{+}$ ions;  this will result in a change in the tunneling spectrum of NH$_{4}^{+}$ ions. The changes in tunneling spectrum of NH$_{4}$ ions in (NH$_{4}$)$_{x}$Rb$_{1-x}$Br  with $x$ (Fig.3)  are, however, partly because of increase in NH$_{4}^{+ }$ - NH$_{4}^{+ }$ interaction and partly because of an increase in the strain field. To examine the role of pure lattice strain on the tunneling spectra of NH$_{4}^{+}$ ion, we need a system where one can vary the lattice strain without altering the NH$_{4}^{+}$ ion concentration. The ternary salts of the type (NH$_{4})_{0.02}($Rb$_{y}$K$_{1-y})_{0.98}$Br or (NH$_{4})_{0.02}($Rb$_{y}$K$_{1-y})_{0.98}$I provide such a system. The lattice strain in above ternary salts can be varied by changing the relative concentration $y$ of Rb$^{+}$ and K$^{+}$ ions without altering the NH$_{4}^{+}$ ion concentration. In the following, results of comparative neutron studies on (NH$_{4})_{0.02}($Rb$_{y}$K$_{1-y})_{0.98}$Br for $y$ = 0.0 (NKB02), 0.5(NKRB02) and 1.0(NRB02) are shown.

\begin{figure}
  \centering
   \includegraphics[width=3.4in,trim=0.09in 0.30in 0.09in 0.13in]{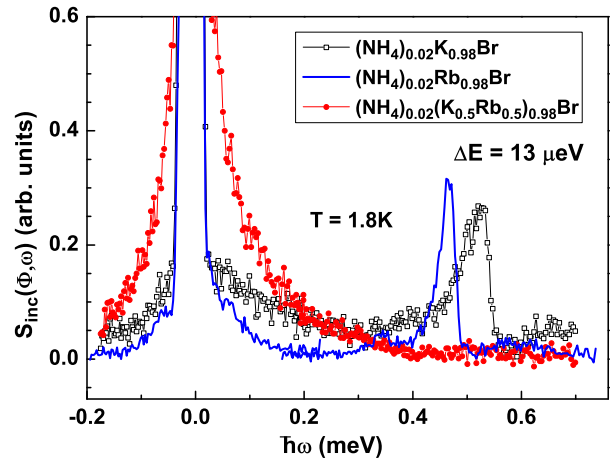}\\
  \caption{Role of strain field on the tunneling spectra of NH$_{4}^{+}$ ions in the mixed salts. The lattice strain in NRKB02 is much larger than that in NRB02 or NKB02.} \label{fig4}
\end{figure}

The measured S$_{inc}(\Phi, \omega $) for  NKB02, NRB02 and NRKB02 are shown in Fig.4. It is seen that while the tunneling spectrum of the NH$_{4}^{+}$ ions in NRB02 or NKB02 show sharp peaks at 0.46 meV and 0.53 meV respectively, the tunneling spectrum for NRKB02 consists of a broad distribution centered around $\hbar \omega $ = 0. As discussed in previous section, the above distribution corresponds to a distribution of tunneling energies in NRKB02. This suggests that NH$_{4}^{+}$ ions experience varying potentials at different sites in NRKB02. The fact that tunneling energies for NH$_{4}^{+}$ ions in NRKB02 are much smaller than those for NRB02 or NKB02, it is reasonable to conclude that rotation of NH$_{4}^{+}$ ions in ternary salt is much more hindered as compared to that in binary salts NRB02 or NKB02.

It is noted that S$_{inc}(\Phi,\omega $) for (NH$_{4})_{0.02}($Rb$_{0.5}$K$_{0.5})_{0.98}$Br (Fig.4) is very similar to the one for (NH$_{4})_{0.44}$K$_{0.56}$I (Fig.2 in reference 6) as measured by Bostoen \textit{et al}.\cite{r6} In both the cases, the tunneling spectra of NH$_{4}^{+}$ ions consists of a distribution of tunneling energies centered about $\hbar\omega $ = 0. This is inspite of the fact that they have widely different concentrations of NH$_{4}^{+}$ ions. The hindering potential V($\theta $) is strong in (NH$_{4}$)$_{0.44}$K$_{0.56}$I because of large NH$_{4}^{+}$ - NH$_{4}^{+}$ interaction, and it is strong in (NH4)$_{0.02}($Rb$_{0.5}$K$_{0.5})_{0.98}$Br due to large contribution from lattice strain. In other words, the  strain field mimics the effect of NH$_{4}^{+}$ - NH$_{4}^{+}$ interaction in  increasing the strength of hindering potential. The dipole freezing temperature, T$_{f}$, depends on the strength of the hindering potential. For example, T$_{f}$ for (NH$_{4}$)$_{x}$K$_{1-x}$I increasing from 4.5 K to 18.0 K as $x$ increases from 0.14 to 0.43.\cite{r5} This suggests that it should be possible to alter the value of T$_{f}$ of a dipolar glass by altering the strain in the lattice. The fact that tunneling energies in (NH$_{4})_{0.02}($Rb$_{0.5}$K$_{0.5})_{0.98}$Br and (NH$_{4})_{0.44}$K$_{0.56}$I are similar, suggests that the hindering potentials and the dipole freezing temperatures are similar ($\sim$ 18 K) for the two salts even when they have widely different concentrations of NH$_{4}^{+}$ ions. In short, these studies would indicate that we can induce dipolar glass transition in a dilute dipolar system by inducing lattice strain.

\section{RESULTS AND DISCUSSIONS: SPECIFIC HEAT STUDIES}

\subsection{Effect of NH$ \mathbf {_{4}^{+}}$ Ion Concentration on Non-Debye Specific Heat of (NH$ \mathbf{ _{4}})_{x}$Rb$_{1-x}$Br}

Figure 5 shows specific heat data for (NH$_{4}$)$_{x}$Rb$_{1-x}$Br for x = 0.0, 0.02, 0.05, 0.10 and 1.0. We have plotted specific heat data as C/T$^{3}$ versus T graphs so that deviations from the Debye law (C $\sim$ T$^{3}$)\cite{r37,r38} are easily seen. The solid lines in the figure correspond to calculated specific heats (see section V for details). At low temperatures, C/T$^{3}$ is nearly constant for pure salts RbBr (x = 0.0) and NH$_{4}$Br (x = 1.0) showing that specific heat of these salts obeys Debye law. It is seen that the specific heat of mixed salts NRB02, NRB05, NRB10 (corresponding to x =0.02, 0.05, 0.10) does not obey Debye law at low temperatures. It may be mentioned that the extra specific heat, referred to as non-Debye specific heat, is not connected with the electronic or magnetic contributions as the mixed salts under discussion are ionic salts and insulators.\cite{r37,r38}

\begin{figure}
  \centering
   \includegraphics[width=3.4in,trim=0.09in 0.30in 0.09in 0.13in]{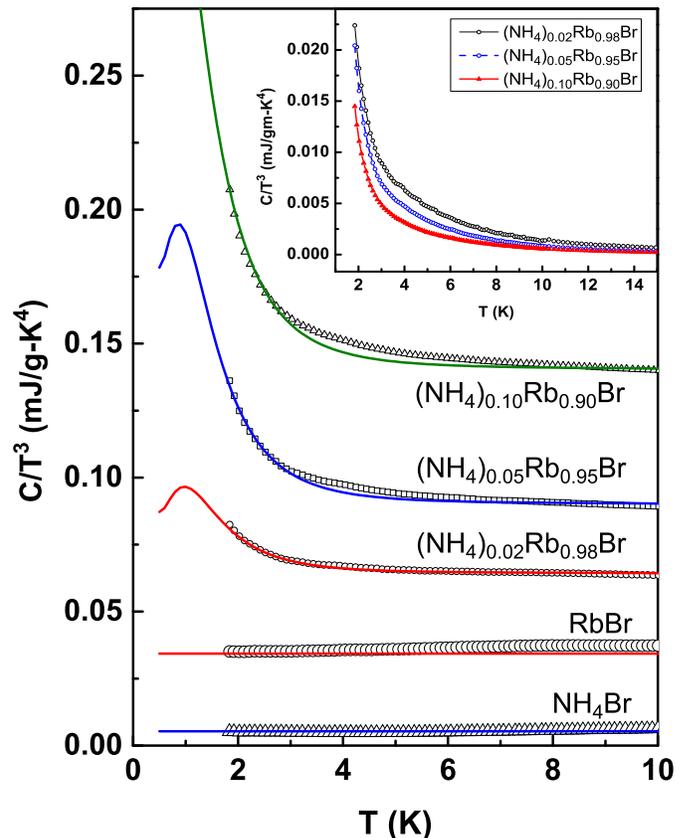}\\
  \caption{Effect of temperature on the specific heat of mixed ((NH$_{4})_{x}$Rb$_{1-x}$Br, x = 0.02, 0.05 and 0.10) and pure salts (NH$_{4}$Br and RbBr). Points are the experimental values and the solid lines are the calculated values (see section V). Inset shows the non-Debye specific heat corrected for the matrix and normalized to a fixed NH$_{4}^{+}$ concentration of $x$ = 0.02.}\label{fig5}
\end{figure}

The specific heat of a crystalline insulating solids such as KBr or NH$_4$Br, is decided by the frequency distribution function g($\omega $) of the normal modes of vibration of the lattice (referred to as phonon density of state) and at low temperatures, it is largely decided by the low energy region of g($\omega $).\cite{r37,r38} As per Debye's theory of specific heat, g($\omega $) for a crystalline material is given by g$_{D}(\omega ) \sim \omega ^{2}$ at low $\omega $ and this leads to the well-known Debye law for specific heat (C$_{v} \sim T^{3}$ ) at low temperatures. However, specific heat of mixed salts depends not only on the Debye phonons but also on the rotational tunneling states of NH$_{4}^{+}$ ions. The tunneling states gives rise to a finite contribution to the specific heat and this is referred to as Schottky specific heat in literature.\cite{r37,r38} The fact that tunneling energies are small ($\sim $ 500 $\mu $eV or $\sim$ 5 K in temperature units), it is expected that the tunneling states would contribute to the specific heat at low temperatures. Usually, specific heat from isolated tunneling states gives rise to a sharp peak (referred to as Schottky peak) in C vs. T graph and the position of the peak depends on the tunneling energy $\Delta E$.\cite{r37} In the present system, there is a distribution of tunneling energies instead of a single tunneling energy and this gives rise to a broadening of Schottky peak. Fig.5 suggests that present data cover a temperature region which corresponds to the tail region of Schottky peak.

It is not clear from Fig.5 if the temperature dependence of non-Debye specific heat is same for all the three NH$_{4}^{+}$ ions concentrations. To get a better insight into the role of dipole concentration on the non-Debye specific heat, the non-Debye component was extracted from the measured specific heat of (NH$_{4})_{x}$Rb$_{1-x}$Br by subtracting the matrix (RbBr) contribution. The non-Debye specific heat curves thus obtained for $x$ = 0.02, 0.05 and 0.10 were normalized to a fixed NH$_{4}^{+}$ concentration of $x$ = 0.02. The normalized non-Debye specific heat curves corresponding to NRB02, NRB05 and NRB10 are shown as an inset in Fig.5. That is, we find that the temperature dependence of non-Debye specific heat of (NH$_{4})_{x}$Rb$_{1-x}$Br changes with $x$. It was seen in the previous section that the rotational tunneling spectra of the NH4 ions in (NH$_{4})_{x}$Rb$_{1-x}$Br also changes with a change in $x$. Thus we conclude that there is a definite correlation between the tunneling spectrum of NH$_{4}^{+}$ ions and the non-Debye specific heat of above salts.

\subsection{Effect of Lattice Strain on Non-Debye Specific Heat of (NH$_{4})_{x}$Rb$_{1-x}$Br}

The effect of lattice strain on the specific heat of (NH$_{4}$)$_{x}$Rb$_{1-x}$Br was examined by carrying out comparative studies on  NRKB02, NRB02 and NKB02. The results are shown in Fig.6. The solid lines in the figure correspond to calculated specific heats (see section V for details).

\begin{figure}
  \centering
   \includegraphics[width=3.4in,trim=0.09in 0.30in 0.09in 0.13in]{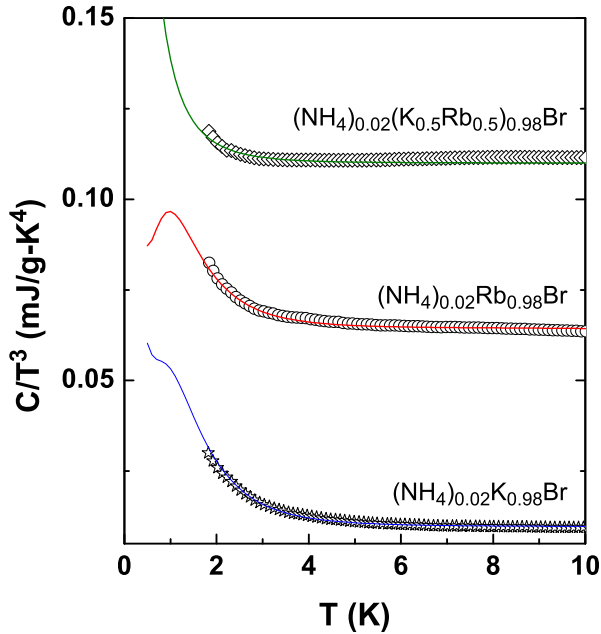}\\
  \caption{Effect of lattice strain on non-Debye specific heat of mixed salt (NH$_{4})_{x}$Rb$_{1-x}$Br. Lattice strain in (NH$_{4})_{0.02}($K$_{0.5}$Rb$_{0.5})_{0.98}$Br is much larger than that in (NH$_4)_{0.02}$Rb$_{0.98}$Br or (NH$_{4})_{0.02}$K$_{0.98}$Br even when all the three samples correspond to a fixed NH$_{4}^{+}$ ion concentration ($x$ = 0.02).}\label{fig6}
\end{figure}

The non-Debye specific heat for NRKB02, NRB02 and NKB02 was obtained by subtracting the matrix [(R$_{0.5}$K$_{0.5}$Br), RbBr and KBr] contribution from the total measured specific heat and is shown in Fig.7. It is noticed that C/T$^{3}$ vs. T curve for NRKB02 is quite different from that of NRB02 or NKB02. In particular, it is seen that the contribution to non-Debye specific heat for NRKB02 starts at lower temperature and increases with decrease in temperature much more rapidly as compared to that for NRB02 or NKB02. That is, present studies clearly show that non-Debye specific heat depends on the strain in the lattice. Watson$^{4}$ had also observed that specific heat depends on the lattice strain in KBr/ KCN/ KCl systems. The above results and the fact that tunneling spectra of NH$_{4}^{+}$ ions in NKRB02 is widely different from that in NKB02 or NKI02 (see section III C), again suggests that there is a definite correlation between the tunneling spectra and the non-Debye specific heat of above salts.

\begin{figure}
  \centering
   \includegraphics[width=3.4in,trim=0.09in 0.30in 0.09in 0.13in]{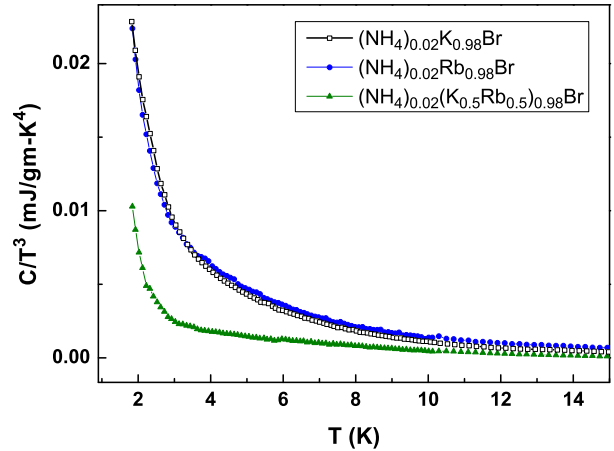}\\
  \caption{Specific heat of binary and ternary salts. While the symbols represent the experimental data, solid lines denote the calculated curves (see section V).}\label{fig7}
\end{figure}

\section{CALCULATION OF SPECIFIC HEAT FROM TUNNELING SPECTRA}

The specific heat of a material, in general, has several different contributions such as phonon, magnetic, electronic and Schottky-type specific heats.\cite{r37,r38} The ammonium and alkali halides and their mixed salts under discussion are ionic salts and their specific heats do not have contributions from electronic or magnetic specific heats. That is,  specific heat of pure salts (e.g. RbBr, KBr, NH$_{4}$Br etc) has contribution from phonons, and the specific heat of mixed salts (e.g. (NH$_{4})_{x}$Rb$_{1-x}$Br, (NH$_{4})_{0.02}($Rb$_{y}$K$_{1-y})_{0.98}$Br etc.) has contributions both from phonons and rotational tunneling states of NH$_{4}^{+}$ ions. It may be mentioned that specific heat of glasses,\cite{r39,r40} including dipolar glasses,\cite{r4} shows a linear variation with temperature at very low temperatures (T $<$ 1 K) which is explained in terms of two level systems (TLS).\cite{r40,r41} The present paper, however, does not deal with TLS or the linear specific heat arising because of TLS. The present specific heat measurements correspond to somewhat higher temperatures (T $>$ 1.5) where non-Debye specific heat is essentially the Schottky specific heat corresponding to the rotational tunneling states of the NH$_{4}^{+}$ ions.

The phonon specific heat C$^{ph}$ of a crystalline material is given by\cite{r37,r38}

\begin{equation}\label{E1}
C^{Ph} = 3R\int _{0}^{\omega_{\max}} \left( \frac{\hbar \omega }{k_{B} T} \right)^{2} \frac{e^{\frac{\hbar \omega }{k_{B} T}} } {\left[ e^{ \frac{\hbar \omega }{ k_{B} T}}  -1 \right]^{2} } \; g(\omega ) d\omega
\end{equation}

\noindent Here g($\omega $) is referred to as phonon density of state and g($\omega ) d\omega $ gives the number of phonons having frequencies between $\omega $ and $\omega  + d\omega $. In case of molecular solids such as ammonium salts, the phonon density of states has contributions from translational (acoustic and optical) modes, rotational modes and internal vibration modes. Depending on their energies, different vibration modes contribute to specific heat in different regions of temperatures\cite{r42}. David Smith\cite{r43} has examined the specific heat for a number of ammonium salts and shown that rotational motion of ammonium ions contribute to the specific heat at T $\sim$ 200K. The specific heat at low temperatures (T $<$ 15 K), the temperature region which is of interest in the present paper, is largely decided by the low energy translational acoustic phonons. Debye argued that the phonon density of state g($\omega $) for a crystalline material is given by  g$_{D}$($\omega $) $\sim$ $\omega ^{2}$ at low $\omega $. The expression for phonon specific heat C$^{Ph}$, relevant to present case, can be written as:\cite{r37,r38}

\begin{equation}\label{E2}
C^{Ph} = \frac{9R}{\omega _{D}^{3} } \int _{0}^{\omega_{D}} \left(  \frac{\hbar \omega }{k_{B} T} \right)^{2} \frac{e^{\frac{\hbar \omega}{k_{B} T}}}{\left[ e^{\frac{\hbar \omega }{k_{B} T}}  - 1 \right]^{2} } \; \omega ^{2} d\omega
\end{equation}

\noindent where cut-off frequency $\omega _{D}$ (= $ \Theta _{D} k_{B} / \hbar $) is decided by the Debye temperature $\Theta _{D}$. As already mentioned, specific heat of pure salts (NH$_{4}$Br, RbBr and KBr) obeys Debye law at low temperatures. The solid lines in Fig.5 corresponding to NH$_{4}$Br and RbBr were calculated using Eq.2 with Debye temperatures of 195 K and 137 K respectively.

The fact that energies ( $\sim$ 500 $\mu$eV, equivalent to 5 K in temperature units)  associated with rotational tunneling states of NH$_{4}^{+}$ ions are small, specific heat of mixed salts at low temperatures depends not only on the Debye phonons but also on the rotational tunneling states of NH$_{4}^{+}$ ions. That is, the specific heat of mixed salts has contributions both from Debye phonons (C$^{Ph}$) and the Schottky specific heat C$^{Sch}$.  It may be mentioned that C$^{Sch}$, which is connected with the rotational tunneling motion of NH$_{4}^{+}$ ions, is different from the rotational specific heat that arises from hindered/free rotation of NH$_{4}^{+}$ ions.\cite{r43} While the rotational tunneling states of NH$_{4}^{+}$ ions contribute to specific heat at low temperatures ( $\sim$ 10 K), the hindered or free rotation of NH$_{4}^{+}$ ions contribute to specific heat at high temperatures ($\sim$ 200 K).\cite{r43} Before we write an expression for C$^{Sch}$ for the present system, it may be mentioned that expression for Schottky specific heat C$^{Sch-2L}$ corresponding to an isolated two level system is given by\cite{r37,r38}

\begin{equation} \label{E3}
C^{Sch-2L} =R\left( \frac{\hbar \omega }{k_{B} T} \right)^{2} \frac{g_{1} }{g_{0} } \frac{e^{-\frac{\hbar \omega}{k_{B} T}} } { \left[ 1 + \frac{g_{1} }{g_{0} } e^{-\frac{\hbar \omega }{k_{B} T}} \right]^{2} }
\end{equation}

\noindent Here $\hbar \omega $ is the tunneling energy, g$_{0}$ is the degeneracy of ground state and g$_{1 }$ is the degeneracy of the excited state. The above simple two level scheme is, however, not valid for describing the tunneling of NH$_{4}^{+}$ ions in mixed salts. The present neutron data show that all NH$_{4}^{+}$ ions are not identical and there is a distribution of tunneling energies. Moreover instead of two level system, the tunneling energy diagram of NH$_{4}^{+}$ ion consists of several energy levels.\cite{r10,r23,r35} Further, the energy level scheme for rotational tunneling states of NH$_{4}^{+}$ ions depends on the symmetry of rotation involved.\cite{r10,r23,r35} For example, energy level diagram and the degeneracy of various levels corresponding to C$_{3V}$ rotation of NH$_{4}^{+}$ ions are different from those for D$_{2d}$ or T$_{d}$ rotations.\cite{r35} In view of above, Eq.3 has been generalized to take account of distribution $f$($\omega $) of tunneling energies. That is, expression for Schottky specific heat is written in terms of $f(\omega $) and then integrated over $\omega $. The modified expression for calculating Schottky specific heat C$^{Sch}$ relevant to the present system is:

\begin{equation} \label{E4}
C^{Sch} =3xR\int _{0}^{\omega _{\max } }{\left( \frac{\hbar \omega } {k_{B} T} \right)^{2} p \frac{e^{-\frac{\hbar \omega }{k_{B} T}}} { \left[ 1 + p \, e^{-\frac{\hbar \omega }{k_{B} T}} \right]^{2} } f(\omega) d\omega}
\end{equation}

\noindent where $x$ is ammonium concentration and $p$ is an effective degeneracy parameter. $f(\omega )$ in Eq.4  is the tunneling density of state, which can be obtained from neutron data (reported in Section III). It can be shown that $f(\omega )$ is given by\cite{r24,r42}

\begin{equation} \label{E5}
f(\omega ) \; \; \sim \; \; \frac{\omega }{Q^{2} } \left( \frac{1}{\coth \left( \frac{\hbar \omega }{2k_{B} T} \right) \pm 1 }
\right) S_{inc} (\Phi ,\omega )
\end{equation}

\noindent S$_{inc}(\Phi, \omega $) in the above expression is the scattering function which was measured in neutron experiment and Q $\left( = E_{i} +E_{f} +2\sqrt{E_{i} E_{f} } \cos (\Phi / 2 )\right)$ is wave vector transfer during scattering. The + and - signs in Eq.(5) correspond to the energy loss and energy gain by the neutrons, respectively. In view of the low temperatures involved, Debye-Waller factor has been assumed to be unity. It may be mentioned that Eq.5 makes use of incoherent approximation for obtaining \textit{f($\omega $}) from S$_{inc}(\Phi, \omega $).\cite{r24} This assumption is quite reasonable because, (i) the measured neutron distributions have been averaged over a range of scattering angles and (ii) the scattering is predominantly from hydrogen of NH$_{4}^{+}$ ions and is incoherent.

The total specific heat for mixed salts where there are contributions from Schottky and Debye specific heats is given by

\setlength{\abovedisplayskip}{0pt} \setlength{\abovedisplayshortskip}{-10pt}
\setlength{\belowdisplayskip}{0pt} \setlength{\belowdisplayshortskip}{3pt}

\begin{equation}\label{E6}
C = C^{Ph} + C^{Sch}
\end{equation}

\noindent In principle, the Debye spectrum and the Debye specific heat C$^{Ph}$ of a mixed salt could change with composition of the salt. We have neglected this effect as the concentration of NH$_{4}^{+}$ ions is small. In any case, the above could increase Debye -- type contribution to specific heat but will not give rise to non-Debye specific heat (say for NRKB02). The solid lines in Figs 5 and 6 corresponding to mixed salts NRB02, NRB05, NRB10, NKB02 and NRKB02 were calculated using equations (4), (5) and (6). In these calculations, we have taken the Debye temperature ($\Theta _{D}$ = $h\omega _{D} /k_{B}$) and the degeneracy parameter $p$ as the fitting parameters. The values of fitted parameters are given in Table-1. The solid lines in Fig.5 corresponding to pure salts NH$_{4}$Br and RbBr are the Debye specific heats for Debye temperatures of 195 K and 137 K respectively. The values of Debye temperatures obtained for the mixed salts are reasonable. That is fitted Debye temperatures for mixed salts NKB02, NRB02, NRB05 and NRB10 are very close to the Debye temperatures of the corresponding pure alkali halides ($\Theta _{D}$ for KBr and RbBr are 173K\cite{r44}  and 137K,\cite{r45} respectively). The fitted Debye temperature for NRKB02 is nearly equal to the average value of the Debye temperatures of KBr and RbBr. In short, it is seen that experimentally measured specific heat of mixed salts of alkali and ammonium halides can be explained in terms of a model that takes account of both Debye phonons and the rotational tunneling states of NH$_{4}^{+}$ ions.

In the present paper, we have reported specific heat, rotational tunneling spectra of NH$_{4}$ ions and a correlation between the two for (NH$_{4}$)$_{x}$Rb$_{1-x}$Br. The general conclusions are, however, expected to be applicable to all dipolar glasses where one is dealing with rotational motion of dipoles. For example, it can be concluded that there is a definite correlation between the rotational tunneling spectra of dipoles and the non-Debye specific heat of dipolar glasses. In general, it should be possible to explain the measured specific heat of a dipolar glass quantitatively in terms of tunneling spectra $f$($\omega $) of dipoles, which is measured independently by a spectroscopic technique.

\noindent \textbf{Table 1:} The values of degeneracy parameter $p$ and the Debye temperature $\Theta _{D}$ as obtained from experimental specific heat data.

\begin{tabular}{|p{1.0in}|p{0.6in}|p{0.6in}|} \hline
\textbf{Sample} & \textbf{p} & ${\mathbf \Theta }_{D}$ \textbf{(K)} \\ \hline
NKB02 & 0.08 & 150 \\ \hline
NRB02 & 0.08 & 118 \\ \hline
NRB05 & 0.08 & 116 \\ \hline
NRB10 & 0.06 & 116 \\ \hline
NKRB02 & 0.03 & 140 \\ \hline
\end{tabular}

\section{SUMMARY}

We have reported the specific heat data, the rotational tunneling spectra of NH$_{4}^{+}$  ions and a correlation between the tunneling states and the specific heat for suitably chosen mixed salts (NRB02, NRB05, NRB10, NKB02, NKI02, and NRKB02) of ammonium and alkali halides, having general formula (NH$_{4}$)$_{x}$A$_{1-x}$Y (A= Rb, K; Y = Br, I). It is seen that measured specific heat of above mixed salts does not obey Debye law and the non-Debye contribution depends on the composition of the salt. We have also measured the rotational tunneling spectra of NH$_{4}^{+}$ ions in above salts in energy ($h\omega $) range of -0.2 meV to +0.70 meV using neutron incoherent inelastic scattering techniques. We have examined the effect of (i) the change in NH$_{4}^{+}$ ion concentration, (ii) the change in first neighbor halide ion of the NH$_{4}^{+}$ ion and (iii) the change in lattice strain  on the on the tunneling spectra $f$($\omega $) and non-Debye specific heat of (NH$_{4}$)$_{x}$Rb$_{1-x}$Br by carrying out measurements on suitably chosen mixed salts.

The role of NH$_{4}^{+}$ concentration has been examined by comparing the results on (NH$_{4}$)$_{x}$Rb$_{1-x}$Br for $x$ = 0.02 (NRB02), 0.05 (NRB05) and 0.10 (NRB10). The tunneling spectrum consists of a strong peak at about 0.46 meV, a small peak at about 0.34 meV and a broad continuous distribution centered around $\hbar \omega $ = 0. The energies of tunneling peaks do not show much change with increase in NH$_{4}^{+}$ ion concentration. The intensity of the broad distribution, however, increases with $x$. The measured non-Debye specific heat also changes with $x$, and agrees with the calculated specific heat from the tunneling spectra.

The effect of first neighbor halide ion was examined by comparing the results on (NH$_{4})_{0.02}$K$_{0.98}$I (NKI02) and (NH$_{4})_{0.02}$K$_{0.98}$Br (NKB02). The general features of tunneling spectrum for the two compounds are nearly same. However, the energy associated with main tunneling peak decreases from 0.606 meV for NKI02 to 0.525 meV for NKB02 and the distribution underneath the elastic peak broadens significantly in NKB02. This shows that the change in halide ion has a significant effect on the tunneling spectra.

The effect of lattice strain has been examined by comparing the results of (NH$_{4})_{0.02}($Rb$_{0.5}$K$_{0.5})_{0.98}$Br (NRKB02), (NH$_{4})_{0.02}$Rb$_{0.98}$Br (NRB02) and (NH$_{4})_{0.02}$K$_{0.98}$Br (NKB02), where the lattice of NRKB02 is much more strained as compared to the other two salts. The sharp tunneling peak seen in NRB02 and NKB02 is absent in NRKB02. However, the distribution under the elastic peak in NRKB02 has not only broadened but also increased in intensity substantially. The present experiments suggest that NH$_{4}^{+}$ ions at different sites in NRKB02 experience varying potentials and this leads to a distribution of tunneling energies in ternary salt. It is concluded that orientational motion of NH$_{4}^{+}$ ions in ternary salt NRKB02 is much more hindered as compared to that in binary salts NRB02 or NKB02.

It is shown that neutron data can be fruitfully used for explaining the non-Debye specific heat of above salts. The specific heat of mixed salts has been analyzed in terms of a model that takes account of contributions from Debye phonons (characterized by Debye temperature, $\Theta_{D}$) and rotational tunneling spectra of NH$_{4}^{+}$ ions. The Schottky specific heat arising from tunneling states was calculated using the rotational tunneling spectra $f$($\omega $) obtained from neutron data. To the best of our knowledge, the present work is a first study where it is shown that measured specific heat of a dipolar glass can be quantitatively explained in terms of an experimentally measured rotational tunneling spectrum $f$($\omega $) of electric dipoles.

\section*{Acknowledgements}

We thank R. Mukhopadhyay for useful discussions and V. Ganesan, R Rawat and S. Pandya for their help in specific heat measurements. We also thank Department of Science and Technology, Govt. of India for providing financial support for carrying out experiments at PSI. One (PSG) of the authors would like to thank UGC-DAE Consortium for Scientific Research for financial support under the project No. CRS-I-34.



\end{document}